\documentclass[lettersize,journal]{IEEEtran}
\IEEEoverridecommandlockouts
\usepackage{amsmath,amssymb,amsfonts}
\usepackage{bbold}
\usepackage{mathtools}
\usepackage[mathscr]{euscript}
\usepackage{graphicx}
\usepackage{textcomp}
\usepackage{xcolor}
\def\BibTeX{{\rm B\kern-.05em{\sc i\kern-.025em b}\kern-.08em
    T\kern-.1667em\lower.7ex\hbox{E}\kern-.125emX}}

\usepackage[style=ieee,backend=biber,bibencoding=utf8,sorting=none,maxbibnames=99]{biblatex}
\bibliography{bibliography}

\usepackage[noend]{algpseudocode}
\usepackage{algorithm}

\usepackage{multirow}
\usepackage{enumitem}
\usepackage{afterpage}

\usepackage[labelformat=parens]{subcaption}

\usepackage{tikz}
\usetikzlibrary{matrix, decorations.pathreplacing, calc, positioning, fit}

\usepackage{blkarray}
\makeatletter
\AtBeginDocument{\BA@colsep=2pt}
\makeatother

\usepackage{cuted}

\usepackage[stable]{footmisc}

\newcommand{\hreflabel}[3][]{%
  \phantomsection
  \label{#2}%
  \href{#1}{#3}%
}

\usepackage[colorlinks=false]{hyperref}
\usepackage{cleveref}

\usepackage{titlesec}
\titlespacing*{\section} {0cm}{0.6cm}{0.0cm}
\titlespacing*{\subsection} {0cm}{0.3cm}{0.2cm}

\begin{document}

\title{An Optimization Algorithm for Customer Topological Paths Identification in Electrical Distribution Networks}

\author{\IEEEauthorblockN{
Maurizio Vassallo\IEEEauthorrefmark{2}\textsuperscript{*},
Adrien Leerschool\IEEEauthorrefmark{3}\textsuperscript{*},
Alireza Bahmanyar\IEEEauthorrefmark{3},
Laurine Duchesne\IEEEauthorrefmark{3},\\
Simon Gerard\IEEEauthorrefmark{4},
Thomas Wehenkel\IEEEauthorrefmark{4},
Damien Ernst\IEEEauthorrefmark{2}
}
\thanks{\textsuperscript{*} Authors contributed equally to the paper.}



\IEEEauthorblockA{
\IEEEauthorrefmark{2}University of Liège,}
\IEEEauthorblockA{\IEEEauthorrefmark{3}Haulogy,}
\IEEEauthorblockA{\IEEEauthorrefmark{4}RESA,}

\IEEEauthorblockA{\{mvassallo, dernst\}@uliege.be}\\
\IEEEauthorblockA{\{adrien.leerschool,
alireza.bahmanyar,
laurine.duchesne\}@haulogy.net}\\
\IEEEauthorblockA{\{simon.gerard, thomas.wehenkel\}@resa.be}
}

\maketitle

\begin{abstract}
A customer topological path represents the sequence of network elements connecting an MV/LV transformer to a customer. Accurate knowledge of these paths is crucial for distribution system operators (DSOs) in digitalization, analysis, and network planning.
This paper introduces an innovative approach to address the challenge of customer topological path identification (TPI) using only the limited and often inaccurate data available to DSOs.
Specifically, our method relies only on geographic information system (GIS) data of network elements and the customer to MV/LV transformers connection information. We introduce an integer linear programming (ILP) optimization algorithm designed to identify customer topological paths that closely approximate the real electricity paths.
The effectiveness of the proposed approach is demonstrated through its application to both an academic and a real-world electrical distribution network. Results show that the method effectively addresses data inaccuracies and successfully identifies customer topological paths, providing a valuable tool for DSOs in developing accurate digital twins of their distribution networks.
\end{abstract}

\section{Introduction}
Power distribution networks are the infrastructure responsible for delivering electricity to consumers.
As distribution networks grow in complexity and scale, driven by factors such as the accommodation of new customers and the increased integration of distributed energy resources (DERs), the need for improved network planning and operation strategies becomes essential. 
A significant challenge in this context is identifying the specific routes electricity takes to reach customers, a task referred to as topological path identification (TPI). Due to the large scale of power distribution networks and limitations of the available data, customer topological paths are generally unknown to the distribution system operators (DSOs).\\
By solving the TPI problem, DSOs gain a clear understanding of customers' connectivity from their houses or businesses to the medium-voltage/low-voltage (MV/LV) substations.
Identifying the topological paths is essential for accurate network digitalization, analysis, and planning. 
\subsection{Literature review}
In power distribution networks, various methodologies have been proposed to address the network topology identification or the TPI problem specifically.
These methods can be characterized into two main categories: methods that only rely on static data such as geographic information system 
(GIS)
data of the elements in the network, and methods that rely on advanced metering infrastructure
(AMI) 
data such as smart sensor recordings.
Methods of the first category generally focus on establishing relationships and connections between elements. Relevant studies include \cite{
RNMLSDP2011,
Seack2014GENERATINGLV,
NavarroEspinosa2015RECONSTRUCTIONOL,
Guzman2017identification,
ZepuKnowledge2020,
li2008distribution,
wu2018distribution,
shang2019automatic,
yin2021research,
wang2023generation}.
For example, in \cite{RNMLSDP2011} and \cite{Seack2014GENERATINGLV} the authors identify the network topology of different networks using publicly available geographical data and connecting the elements together based on their distances.
The authors in \cite{NavarroEspinosa2015RECONSTRUCTIONOL} focus on network topology identification using Euclidean distances and breadth-first search to identify clusters in the network, followed by refining the topology by connecting nodes to the most probable lines.
In \cite{Guzman2017identification}, the authors present a kd-tree algorithm for identifying clusters of elements, followed by connecting the elements within each cluster based on a predefined distance.
The paper \cite{ZepuKnowledge2020} uses knowledge graphs to reflect the existing relationship among the elements in low-voltage distribution networks. The knowledge graphs are then used to identify the correct network topology.
\\
The main advantage of these methods is that they operate without relying on AMI measurements, making them easier to implement for DSOs with limited advanced infrastructure. However, their major limitation is that GIS data can be incomplete, inaccurate, or missing for certain areas of the network. \\
The second category of solutions relies on AMI measurements from the network. Relevant studies include \cite{
farajollahi_topology_2020,
RLVNLO2021,
soltani_real-time_2022,
WangPower2021,
parikh2009transforming,
peppanen2016estimation,
morrell2018modelling,
cui2022low,
li2022topology,
Yu2024identification,
PTI2024,
benzerga2021low,
}. For example,
in \cite{farajollahi_topology_2020}, the authors use line current sensor measurements and a mixed-integer linear programming (MILP) optimization algorithm to identify the topology of a power distribution network in California. 
The authors in \cite{RLVNLO2021} introduce an iterative methodology for reconstructing network topology and cable parameters of low-voltage three-phase networks with limited smart meter coverage.
The authors in the paper \cite{soltani_real-time_2022} propose a similar approach using mixed-integer quadratic programming (MIQP) to identify the topology of power distribution networks using smart sensor recordings.
This category of methods can estimate the topology and often can estimate line parameters (e.g., impedance)  as well. However, they often require AMI measurements at each customer connection point. This limitation restricts their applicability in networks with incomplete AMI coverage, as evidenced by some studies using limited smart meter data \cite{benzerga2021low}.

\subsection{Paper contribution}
This paper presents a novel approach for assisting DSOs in addressing the TPI problem using only the available static data, even if this data is often incomplete and inaccurate. The proposed method relies exclusively on GIS data of network elements and the customers' connections to MV/LV transformers, without requiring any AMI measurements.
The problem is formulated as an integer linear programming (ILP) optimization algorithm.
The objective function of the optimization problem focuses on maximizing the number of customers connected to the correct MV/LV transformer, while some constraints ensure that the solution satisfies the DSO's expectations.
The proposed methodology effectively addresses data inaccuracies and incompleteness.\\
Moreover, a repository with the code of the proposed methodology is released. The repository is publicly available on GitHub at the following link: \hreflabel[https://github.com/TPIproblem/OptimalTPI]{repo}{https://github.com/TPIproblem/OptimalTPI}.

\subsection{Paper organization}
\noindent The rest of the paper is organized as follows:
Section \ref{nettop} presents the definition of the power network elements considered in this work.
Section \ref{tp} defines the concept of topological paths.
The problem statement is presented in Section \ref{ps}.
Section \ref{methodology} defines the methodology used to solve the problem.
The methodology is applied to an academic example in Section \ref{accexample} and a real Belgian power distribution network in Section \ref{casestudy}.
Section \ref{conclusion} concludes the work and discusses possible future works. 

\section{Power distribution modeling
\footnote{The definitions used in this section were first discussed in the work \cite{TPI2024}.}} \label{nettop}

\noindent Power distribution networks typically encompass a wide range of elements.
\noindent The set of elements $\mathcal{E}$ is defined as:
\begin{align}
    \mathcal{E} = \{e_1, ..., e_k, ..., e_{|\mathcal{E}|}\}
\end{align}
\noindent where $|\mathcal{E}|$ represents the total number of elements in the set $\mathcal{E}$ and a single element is denoted as $e_k$, with $k \in \{1, ..., |\mathcal{E}|\}$.\\


\noindent Each element can have some attributes, such as its type and coordinates. The set of all element attributes $\mathcal{A}$ is defined as follows:
\begin{align}
    \mathcal{A} = \{a_1, ..., a_{|\mathcal{A}|}\}.
\end{align}

\noindent A single element $e \in \mathcal{E}$ is defined as a set of attributes.
Formally:
\begin{align}
    e = \{ e.a_1, ..., e.{a_k} ..., e.a_{|e|} \}
\end{align}
where the dot notation is used to access the attributes of the element $e$, and $a_k$ represents, for example, the type of the element $e$, its coordinates, or other relevant properties.\\


\noindent Each element within the network can be categorized based on its type attribute, serving as a characteristic that distinguishes it from other elements. The set of types, denoted as $\mathcal{T}$, includes all the possible types associated with the elements in $\mathcal{E}$:
\begin{align}
    \mathcal{T} = \{t_1, ..., t_{|\mathcal{T}|}\}.
\end{align}
\noindent Among the typical types, there may be $customer$ connection elements, MV/LV $transformer$ substations.\\
Given an element $e \in \mathcal{E}$, its type, $t \in \mathcal{T}$, is accessed as follows:
\begin{align}
    t = e.type.
\end{align}
\subsection{Network topological functions}
\noindent The $Subset()$ function is used to identify all elements of a specific type. This function takes two input parameters: the element set, $\mathcal{E}$, and a type $t \in \mathcal{T}$. 
Formally: 
\begin{align}
    Subset(\mathcal{E}, t) =  \{ e \in \mathcal{E} \; | \; e.type = t \}.
\end{align}
\noindent In particular, three useful subsets are defined:
\begin{itemize}
    \item[$\mathtt{C}$:] represents the set of customers in the network, $\mathtt{C} = Subset(\mathcal{E}, customer)$.
    \item[$\mathtt{T}$:] represents the set of terminal elements, generally $\mathtt{T} = Subset(\mathcal{E}, transformer)$. Note that the terminal element can vary depending on the context. For example, in this paper, we consider the terminal element to be the feeder junction where the feeder starts in the MV/LV substation, but it could also be extended to the high-voltage/MV substation or any other element in the network. The methodology is designed to be flexible.
    \item[$\mathtt{R}$:] represents the set of all elements, excluding customers and terminal elements, $\mathtt{R} = \mathcal{E} - \mathtt{C} - \mathtt{T}$.\\
\end{itemize}

\noindent The $Dist()$ function takes two elements $e_k, e_m \in \mathcal{E}$ as input, and it outputs a scalar value, $d$, representing the distance between them.
The $Dist()$ function is defined as follows:
\begin{align}
\label{eq:dist}
    d &= Dist(e_k,e_m) = ||e_k.coor, \, e_m.coor||_2
\end{align}

\noindent where $||\bullet||_2$ represents the Euclidean distance and where $e_k.coor$ and $e_m.coor$ give the coordinates of the element $e_k$ and, $e_m$ respectively.\\


\noindent The $Connections()$ function identifies all elements in the set $\mathcal{E}$ that can be connected to a given element, $e \in \mathcal{E}$, considering some specific conditions. The conditions for connectivity can depend on the DSOs' requirements such as distance between the elements, the element types, and so on. \\
Therefore, the $Connections()$ function takes three inputs: $\mathcal{E}$, $e$ and some conditions, returning a subset $\mathtt{K} \subset \mathcal{E}$. Formally:
\begin{align}
\label{eq:connections}
    \mathtt{K} = \; & 
    Connections(\mathcal{E} , e, conditions) = \\
     & \{e_m \in \mathcal{E} \; | \; e_m \text{ is an element that can be directly connected} \nonumber \\ 
    &\;\;\;\;\;\;\;\;\;\;\;\;\;\;\; \text{to $e$ if they satisfy the given conditions}\}. \nonumber
\end{align}


\section{Topological paths} \label{tp}
This section serves as the foundation for understanding the concepts of hypothetical, real, and estimated paths within the context of this work.
\subsection{Paths}
A customer topological path is represented as an ordered list of elements:
\begin{align}
    p = (p_1, p_2, ..., p_k, ..., p_{|p|-1}, p_{|p|})
\end{align}
where $p$ is a generic path, $p_1$ represents the first element in the path which is a customer; $p_k$ represents the $k$-element in the path, and $p_{|p|}$ represents the terminal element, generally an MV/LV transformer.\\
Therefore, $p_1 \in \mathtt{C}$, and $p_k \in \mathtt{R}$ with $k \in \{2, ..., |p|-1\}$ and $p_{|p|} \in \mathtt{T}$.

\subsection{Hypothetical paths}
Hypothetical paths are paths where only the initial elements are known with certainty, while the connections between intermediate elements up to the terminal point are unknown. These paths represent potential routes that electricity might follow to supply a customer.\\ 
\noindent The set of hypothetical paths, denoted as $\mathcal{H}$.
Each individual hypothetical path, $h \in \mathcal{H}$, represents a hypothetical path that electricity can follow to supply a customer.
The number of hypothetical paths, given the set of elements $\mathcal{E}$, is finite, and it is given by Eq. (18) in paper \cite{TPI2024} and here reported:
\newcommand*{\Perm}[2]{{}^{#1}\!P_{#2}}%
\begin{align} \label{eq:size_h}
    |\mathcal{H}| = |\mathtt{C}| \cdot \sum_{k=1}^{|\mathtt{R}|} \!\! \Perm{|\mathtt{R}|}{k} \cdot |\mathtt{T}|
\end{align}
where $\Perm{n}{m}$ is the permutation formula, which calculates the number of ways to arrange $k$ items from a set of $n$ items ($\Perm{n}{m} = \frac{n!}{(n-m)!}$).

\subsection{Real paths} 
The set of all real paths, denoted as $\mathcal{P}$, is a subset of the hypothetical path set $\mathcal{H}$. Each path $p \in \mathcal{P}$ uniquely represents the actual route that electricity takes from an MV/LV transformer to a customer.

\noindent In radial networks, the number of real paths is equal to the number of customers, since one customer is supplied by only one MV/LV transformer at any given moment. Therefore, the following condition holds $|\mathcal{P}| = |\mathtt{C}|$,
where $\mathtt{C}$ represents the set of elements of type customers.

\subsection{Estimated paths}
The estimated paths, denoted by $\hat{\mathcal{P}}$, form a subset of the hypothetical paths, $\mathcal{H}$. This set of estimated paths, $\hat{\mathcal{P}}$, is an approximation to the set of real paths within the network, $\mathcal{P}$.

\noindent The set $\hat{\mathcal{P}}$ holds the property that the number of estimated paths in the set is equal to the number of customers considered in the network. Therefore, the following condition holds $|\hat{\mathcal{P}}| = |\mathcal{P}| = |\mathtt{C}|$.



\subsection{Paths visualization}
\begin{figure}[h!]
\hfill
\centering
\begin{subfigure}[b]{0.4\textwidth}
        \centering
        \includegraphics[width=1\textwidth]{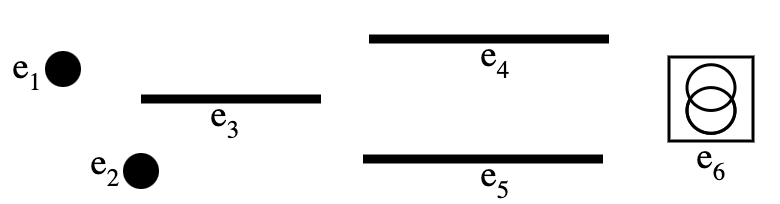}
        \caption{Simple network}
        \label{fig:pathvisualizationA}
\end{subfigure}
\hfill
\vspace{2mm}
\begin{subfigure}[b]{0.4\textwidth}
        \centering
        \includegraphics[width=1\textwidth]{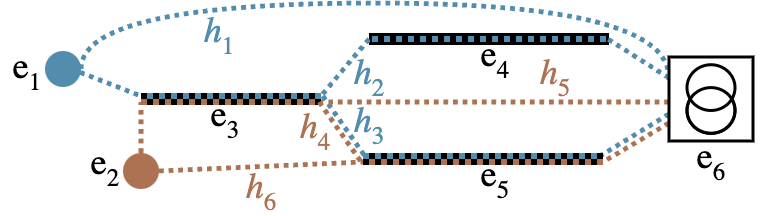}
        \caption{Hypothetical paths}
        \label{fig:pathvisualizationB}
\end{subfigure}
\hfill
\begin{subfigure}[b]{0.4\textwidth}
        \centering
        \includegraphics[width=1\textwidth]{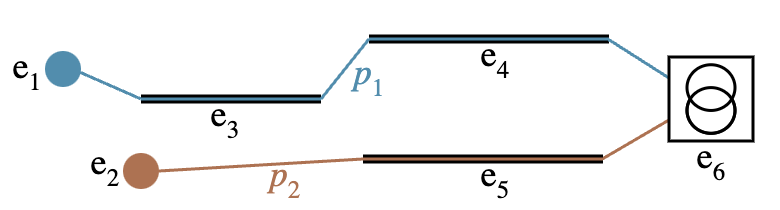}
        \caption{Real paths}
        \label{fig:pathvisualizationC}
\end{subfigure}
\hfill
\begin{subfigure}[b]{0.4\textwidth}
        \centering
        \includegraphics[width=1\textwidth]{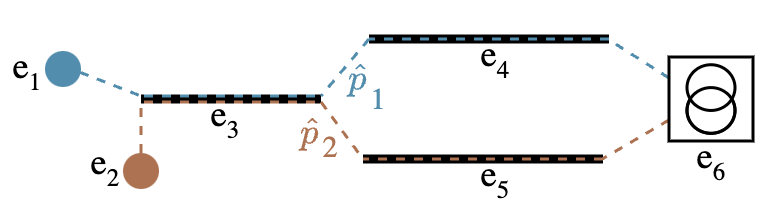}
        \caption{Estimated paths}
        \label{fig:pathvisualizationD}
\end{subfigure}
\hfill
\caption{Visualization of the different sets of paths.}
\label{fig:pathvisualization}
\end{figure}

Figure \ref{fig:pathvisualization} illustrates a simplified network, highlighting hypothetical, real, and estimated paths.

\noindent As shown in Fig. \ref{fig:pathvisualizationA} the data about network elements are often incomplete, making it difficult to accurately identify the customer paths.\\
Figure \ref{fig:pathvisualizationB} shows some of all hypothetical paths. In the absence of prior information, any path could potentially represent the real one. 
This highlights the challenge that with limited information, only limited conclusions can be drawn about whether a hypothetical path corresponds to its real one. Many hypothetical paths for each customer may be available.\\
Figures \ref{fig:pathvisualizationC} and \ref{fig:pathvisualizationD} illustrate the real and estimated paths for each customer, respectively. Each customer is associated with exactly one real path and one estimated path, both of which are subsets of the hypothetical path set. Therefore: $\mathcal{P} \subset \mathcal{H}$ and $\hat{\mathcal{P}} \subset \mathcal{H}$. 

\begin{figure}[h!]
    \centering
    \includegraphics[width=0.3\textwidth]{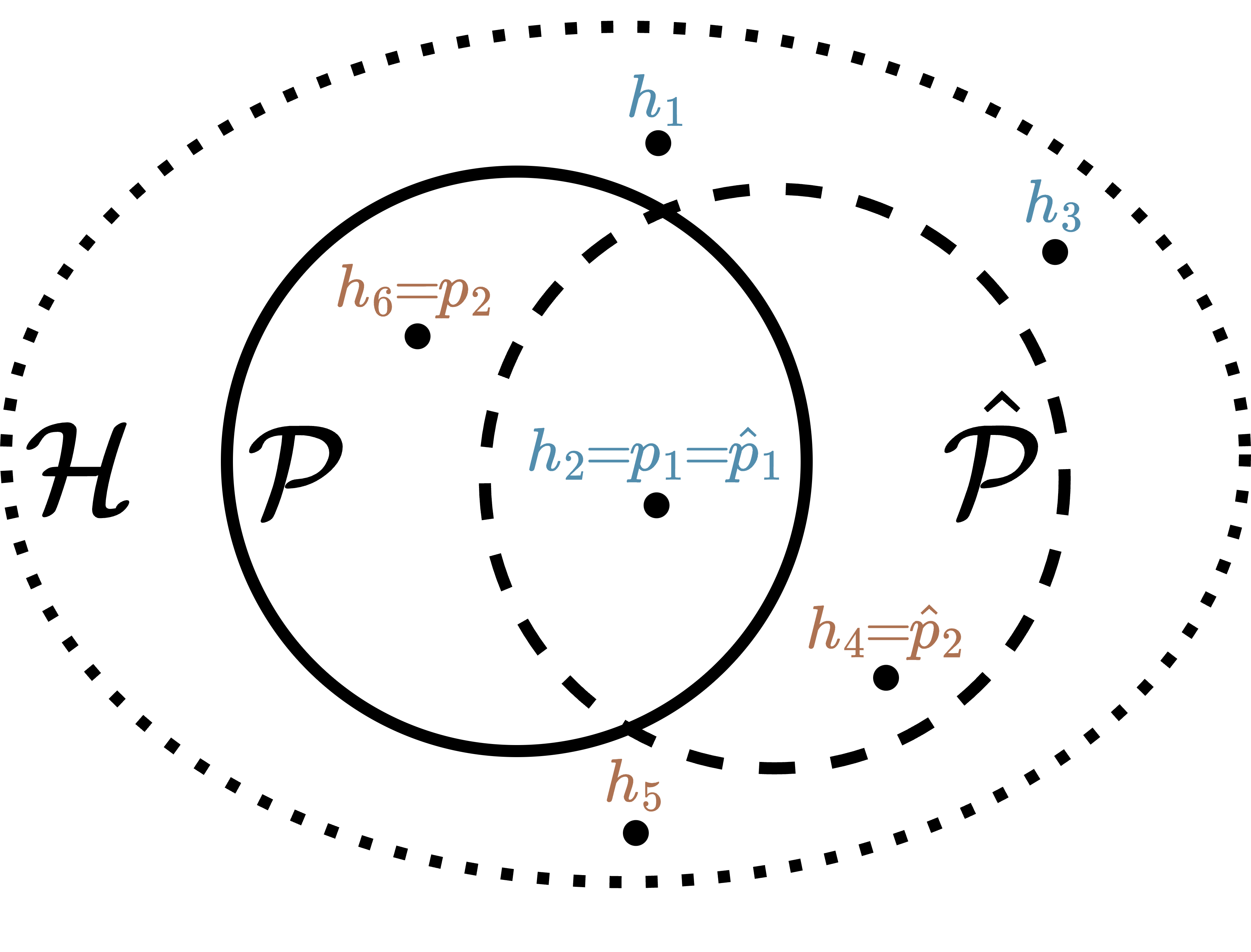}
    \caption{Representation of the different sets and their relationships.}
    \label{fig:pathsets}
\end{figure}
\noindent 
Figure \ref{fig:pathsets} presents a Venn diagram illustrating the relationships among the path sets. The set of hypothetical paths, $\mathcal{H}$, encompasses all other sets, representing all possible paths within the network. Notably, the sets of real paths, $\mathcal{P}$, and estimated paths, $\hat{\mathcal{P}}$, overlap to a certain degree, highlighting similarities and discrepancies between the estimation and ground truth. \noindent The overlap between $\mathcal{P}$ and $\hat{\mathcal{P}}$, indicates that some estimated paths match the real ones while some others do not. For example, the paths $\hat{p}_1$ and $p_1$ for customer $e_1$, match perfectly. However, data inaccuracies can result in discrepancies, for example, the estimated path $\hat{p}_2$ deviates from the real path $p_2$ for the customer $e_2$.

\section{Problem statement} \label{ps}


Generally, DSOs possess different types of information. The information available to the DSO is denoted as $\mathcal{I}$. This data can include, for example, files about the GIS data of the network elements, customers' information like annual power consumption, network configuration rules explaining how the elements are connected, and more.\\
The goal of the DSOs is, given all the raw information available to them, to estimate the customer topological paths that are as close as possible to the real paths. Denoting the set of real paths as the set $\mathcal{P}$ and the set of estimated paths as $\hat{\mathcal{P}}$, the objective is to identify two sets to be as close as possible and in the best case equal 
($\hat{\mathcal{P}} = \mathcal{P}$).
\\
Therefore, the problem is to identify a methodology, $\mathfrak{M}()$, that, starting from the set of raw information, $\mathcal{I}$ can identify the best approximation of the real paths $\mathcal{P}$.
Formally:
\begin{align}
\hat{\mathcal{P}} = \mathfrak{M}(\mathcal{I}) \quad \text{s.t.} \quad \hat{\mathcal{P}} \simeq \mathcal{P}.
\end{align}

\section{Methodology} \label{methodology}
\begin{figure}[h!]
    \centering
    \includegraphics[width=0.49\textwidth]{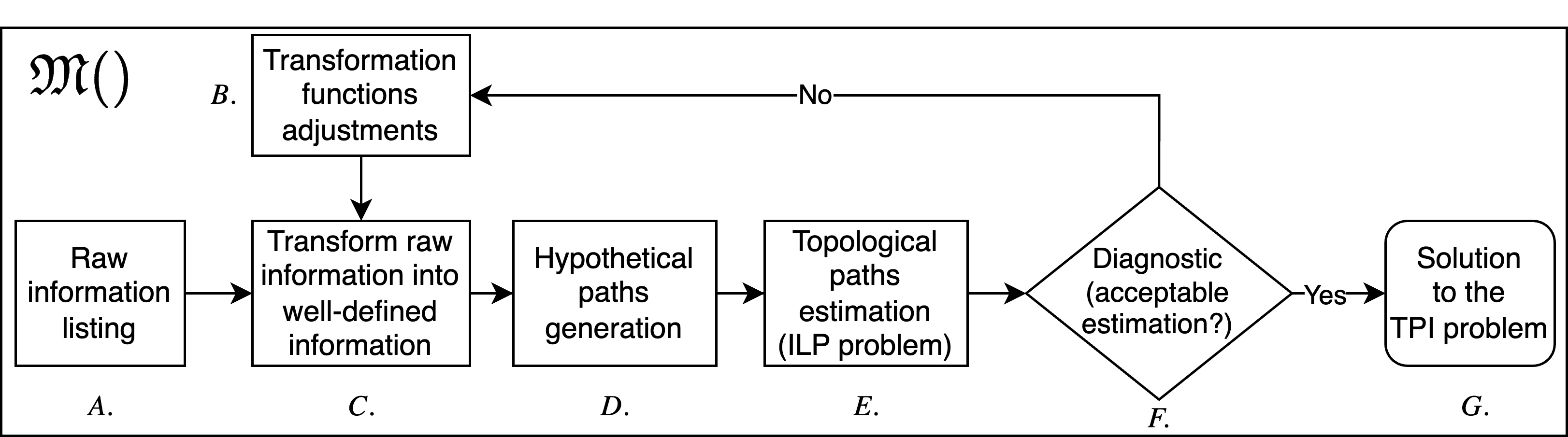}
    \caption{Flowchart of the steps proposed by our methodology, $\mathfrak{M}()$, to identify the customer topological paths in electrical distribution networks.}
    \label{fig:methodology}
\end{figure}

This section outlines the step-by-step process of the proposed methodology, 
$\mathfrak{M}()$, designed to identify customer topological paths in electrical distribution networks. A visual representation of the methodology is provided in Fig. \ref{fig:methodology}.
\subsection{Raw information listing}
Raw information refers to available information or data that come from various sources. 
The set of raw information is listed as follows: $\mathcal{I} = \{i_1, ..., i_k, ..., i_{|\mathcal{I}|} \}$, with $i_k$ ($k \in \{1, ..., |\mathcal{I}|\}$) denoting a single piece of raw information.\\

\noindent For this paper, we use only static data, and in particular, the data needed is the GIS coordinates of the different elements in the network, their type, and the feeder terminal junction to which each customer is connected. Generally, this kind of information is available to the DSOs, even if it may be incomplete and/or inaccurate.\\
Formally, the data needed for this paper is:
\begin{itemize}
    \item The set of elements, $\mathcal{E}$.
    \item The set of attributes, $\mathcal{A}$, contains at least the coordinates of the elements, the types, and the feeder terminal junction to which each customer is connected.
    \item The set of types $\mathcal{T}$, with $\mathcal{T}$ containing at least customer, line, junction, and transformer.
    \item $e.coor$, $e.type$ are known for all the elements, $e \in \mathcal{E}$.
    \item $c.junction$ is known for all the customers, $c \in \mathtt{C}$.
\end{itemize}

\subsection{Transformation functions adjustments}
Transformation functions are responsible for transforming the available raw information into clear knowledge that is relevant within the specific context of a power network.
The set of transformation functions is denoted as follows:
$\mathcal{F} =\{ f_1, ..., f_m, ..., f_{|\mathcal{F}|} \}$, with $f_m$ ($m \in \{1, ..., |\mathcal{F}|\}$) denoting a single transformation function. \\
Mathematically, the transformation functions take one piece of raw information as input and return the transformed information as output.

\subsection{Well-defined information}
The well-defined information forms the basis to provide a structured foundation to identify customer paths that are as close as possible to reality.
The set of well-defined information is denoted as follows: $\mathcal{I}' =\{ i'_1, ..., i'_n, ..., i'_{|\mathcal{I}'|} \}$, with $i'_n$ ($n \in \{1, ..., |\mathcal{I}'|\}$) denoting a single piece of well-defined information. Each element in the set $\mathcal{I}'$ is the result of applying a transformation function on a piece of raw information, i.e. $i'_n = f_m(i_k)$.

\subsection{Hypothetical paths generation}
Given the set of well-defined information, $\mathcal{I}'$, the set of hypothetical path compatible with the set $\mathcal{I}'$, denoted as $\mathcal{H}^{\mathcal{I}'}$, is constructed.
Each piece of information in $\mathcal{I}'$ allows to reduce the size of the hypothetical path set $\mathcal{H}^{\mathcal{I}'}$.\\
The idea is that the more data is available, the closer to reality the paths inside the set $\mathcal{H}^{\mathcal{I}'}$ are.

\subsection{Topological paths estimation}
After generating the hypothetical paths compatible with the well-defined information, $\mathcal{H}^{\mathcal{I}'}$, the next step of the methodology is to identify the paths in the set $\mathcal{H}^{\mathcal{I}'}$ that are an approximation of the real paths in the actual network, therefore identifying the set of estimated paths $\hat{\mathcal{P}}$.\\
While various methods can solve the TPI problem, we present an ILP approach designed to find the solution $\hat{\mathcal{P}}$ and to meet the DSO requirements.

\subsubsection{Matrices generation}
\label{ss:mat_dec}
The following sections describe the main components of our ILP formulation, including the matrices and the optimization problem.
\begin{itemize}[leftmargin=3mm]
    \item Hypothetical paths matrix:\\
        The set of hypothetical paths, $\mathcal{H}^{\mathcal{I}'}$ can be equivalently interpreted as a binary matrix with a number of rows equivalent to $|\mathcal{H}^{\mathcal{I}'}|$ and a number of columns equivalent to $|\mathtt{C}|+|\mathtt{R}|+|\mathtt{T}|$.\\
   Therefore, the hypothetical paths matrix, denoted as $\mathbf{H}$, can be written as follows:
   \definecolor{grenMatrix}{rgb}{0.0, 0.5, 0.0}
\begin{equation}
\label{eq:matrixH}
 \!\!\!\!\! \mathbf{H} \! = \!\!\!\!\!\!\!\!\!\!
\begin{tikzpicture}[
    baseline,
    label distance=0pt, 
    every right delimiter/.style={xshift=-.4em}, 
    every left delimiter/.style={xshift=.4em},
]

\matrix [matrix of math nodes,left delimiter=(,right delimiter=),row sep=0.1cm,column sep=0.053cm] (m) {
    0 & 1 & \cdots & 0 & 1 & 0 & \cdots & 1 & 0 & 1 & \cdots & 0 \\
    1 & 0 & \cdots & 0 & 1 & 0 & \cdots & 0 & 0 & 1 & \cdots & 0 \\
    \vdots & \vdots & \ddots & \vdots &\vdots & \vdots & \ddots & \vdots &\vdots & \vdots & \ddots & \vdots \\
    0 & 0 & \cdots & 0 & 0 & 0 & \cdots & 0 & 0 & 0 & \cdots & 1 \\
    \vdots & \vdots & \ddots & \vdots &\vdots & \vdots & \ddots & \vdots &\vdots & \vdots & \ddots & \vdots \\
    0 & 0 & \cdots & 1 & 0 & 1 & \cdots & 0 & 1 & 0 & \cdots & 0 \\
};

\node[
  fit=(m-1-1),
  inner ysep=1mm,
  label=above:$c_1$
] {};
\node[
  fit=(m-1-2),
  inner ysep=1mm,
  label=above:$c_2$
] {};
\node[
  fit=(m-1-3),
  inner ysep=2mm,
  label=above:$\cdots$
] {};
\node[
  fit=(m-1-4),
  inner ysep=.2mm,
  label=above:$c_{|\mathtt{C}|}$
] {};
\node[
  fit=(m-1-5),
  inner ysep=1mm,
  label=above:$r_1$
] {};
\node[
  fit=(m-1-6),
  inner ysep=1mm,
  label=above:$r_2$
] {};
\node[
  fit=(m-1-7),
  inner ysep=2mm,
  label=above:$\cdots$
] {};
\node[
  fit=(m-1-8),
  inner ysep=0.2mm,
  label=above:$r_{|\mathtt{R}|}$
] {};
\node[
  fit=(m-1-9),
  inner ysep=1mm,
  label=above:$t_1$
] {};
\node[
  fit=(m-1-10),
  inner ysep=1mm,
  label=above:$t_2$
] {};
\node[
  fit=(m-1-11),
  inner ysep=2mm,
  label=above:$\cdots$
] {};
\node[
  fit=(m-1-12),
  inner ysep=.2mm,
  label=above:$t_{|\mathtt{T}|}$
] {};

\node[
  fit=(m-1-1),
  inner xsep=6pt,inner ysep=0,
  label={[label distance=0pt]left:$h_1$}
] {};

\node[
  fit=(m-2-1),
  inner xsep=6pt,inner ysep=0,
  label={[label distance=0pt]left:$h_2$}
] {};
\node[
  fit=(m-3-1),
  inner xsep=10pt,inner ysep=0,
  label={[label distance=0pt]left:$\vdots$}
] {};
\node[
  fit=(m-4-1),
  inner xsep=6pt,inner ysep=0,
  label={[label distance=0pt]left:$h_k$}
] {};
\node[
  fit=(m-5-1),
  inner xsep=10pt,inner ysep=0,
  label={[label distance=0pt]left:$\vdots$}
] {};
\node[
  fit=(m-6-1),
  inner xsep=2pt,inner ysep=0,
  label={[label distance=0pt]left:$h_{|\mathcal{H}^{\mathcal{I}'}|}$}
] {};

\draw[dashed, red] ([xshift=1.5pt] m-1-1.north west) rectangle ([xshift=-1.5pt] m-6-4.south east) node[midway, below=58pt] {$\mathbf{H}_{\mathtt{C}}$};

\draw[dashed, grenMatrix] ([xshift=1.5pt] m-1-5.north west) rectangle ([xshift=-1.5pt] m-6-8.south east) node[midway, below=58pt] {$\mathbf{H}_{\mathtt{R}}$};

\draw[dashed, blue] ([xshift=1.5pt] m-1-9.north west) rectangle ([xshift=-1.5pt] m-6-12.south east) node[midway, below=58pt] {$\mathbf{H}_{\mathtt{T}}$};

\end{tikzpicture} 
\end{equation}
    
    \hspace{-0.52em} Each row $h_k$ in the matrix $\mathbf{H}$ represents a hypothetical path compatible with the well-defined information. A value of $1$ or $0$ in this row indicates whether the corresponding element, $e \in \mathtt{C} \cup \mathtt{R} \cup \mathtt{T}$, is present ($1$) or not ($0$) in path $h_k$. A single element of the matrix is accessed as $\mathbf{H}_{(k,\,m)}$ where $m$ represents the row ($k \in \{1, ..., |\mathcal{H}^{\mathcal{I}'}|\}$) and $m$ represents the column ($m \in \{1, ...,  |\mathtt{C}| + |\mathtt{R}| + |\mathtt{T}|\}$).\\
   
   \noindent For clarity, we divide the $\mathbf{H}$ matrix into three sub-matrices:
   \begin{itemize}
       \item The $\mathbf{H}_{\mathtt{C}}$ matrix, denoted as customers elements matrix, indicates the presence of a customer $c \in \mathtt{C}$ in any path, $h \in \mathcal{H}^{\mathcal{I}'}$. Therefore, the $\mathbf{H}_{\mathtt{C}}$ matrix has dimension $|\mathcal{H}^{\mathcal{I}'}|\times|\mathtt{C}|$. \\
       Since one and only one customer can be present in one path, the following condition must be satisfied:
       \begin{align}
           \sum_{k=1}^{|\mathtt{C}|} \mathbf{H}_{\mathtt{C}\,(m,\,k)}=1, \forall m \in \{1, ..., |\mathcal{H}^{\mathcal{I}'}|\}.
       \end{align}
       \item The $\mathbf{H}_{\mathtt{R}}$ matrix, denoted as remaining elements matrix, indicates the presence of elements $r \in \mathtt{R}$ in the paths, $h \in \mathcal{H}^{\mathcal{I}'}$. Therefore, the $\mathbf{H}_{\mathtt{R}}$ matrix has dimension $|\mathcal{H}^{\mathcal{I}'}|$$ \,\times\,|\mathtt{R}|$.\\
       No particular condition is imposed on the matrix $\mathbf{H}_{\mathtt{R}}$.
       
       \item The $\mathbf{H}_{\mathtt{T}}$ matrix, referred as terminal elements matrix, indicates the presence of a terminal node, $t \in \mathtt{T}$, in the paths, $h \in \mathcal{H}^{\mathcal{I}'}$. Therefore, the $\mathbf{H}_{\mathtt{T}}$ matrix has dimension $|\mathcal{H^{\mathcal{I}'}}|\times|\mathtt{T}|$. \\
       Since one and only one terminal element can be present in one path, the following condition must be satisfied:
       \begin{align}
       \label{eq:h_trafo}
           \sum_{k=1}^{|\mathtt{T}|} \mathbf{H}_{\mathtt{T}\,(m,\,k)}=1, \forall m \in \{1, ..., |\mathcal{H}^{\mathcal{I}'}|\}.
       \end{align}
   \end{itemize}

    \item Terminal association matrix:\\
        We define a binary matrix, referred to as the terminal association matrix, denoted as $\mathbf{T}_{\mathtt{R}}$, which indicates the association of elements $r \in \mathtt{R}$ to a specific terminal element $t \in \mathtt{T}$.\\
        The $\mathbf{T}_{\mathtt{R}}$ matrix has dimension $|\mathtt{T}|\times|\mathtt{R}|$ and can be represented as follows:
        \begin{align}
            \mathbf{T}_{\mathtt{R}} = \begin{blockarray}{cccccccc}
                  & r_1 & r_2 & \cdots & r_m & \cdots & r_{|\mathtt{R}|-1} & r_{|\mathtt{R}|}\\
            \begin{block}{c(ccccccc)}
              t_1               \;\; & 0 & 1 & \cdots & 0 & \cdots & 0 & 0  \\
              t_2               \;\; & 1 & 0 & \cdots & 0 & \cdots & 1 & 0  \\
              \vdots            \;\; & \vdots & \vdots & \ddots & \vdots & \ddots & \vdots & \vdots \\
              t_k               \;\; & 0 & 0 & \cdots & 1 & \cdots & 0 & 0  \\
              \vdots            \;\; & \vdots & \vdots & \ddots & \vdots & \ddots & \vdots & \vdots \\
              t_{|\mathtt{T}|}  \;\; & 0 & 0 & \cdots & 0 & \cdots & 0 & 1  \\
            \end{block}
          \end{blockarray}
        \end{align}
        where $0$ and $1$ represent respectively whether an element $r \in \mathtt{R}$ is assigned to a terminal element $t \in \mathtt{T}$ or not.\\

    \item Estimated paths matrix:\\
            The solution of the TPI problem, the set $\hat{\mathcal{P}}$, can be also conceptualized as a binary matrix that represents the hypothetical paths estimated to be the real paths. This path matrix, denoted as $\hat{\mathbf{P}}$, has dimension $1 \times |\mathcal{H}^{\mathcal{I}'}|$.
            \begin{align}
                \hat{\mathbf{P}} = \begin{blockarray}{ccccc}
                  h_1 & \cdots & h_k & \cdots & h_{|\mathcal{H}^{\mathcal{I}'}|} \\
                \begin{block}{(ccccc)}
                  1 & \cdots & 1 & \cdots & 0\\
                \end{block}
              \end{blockarray}
            \end{align}
            where $0$ and $1$ represent whether the path $h_k \in \mathcal{H}^{\mathcal{I}'}$ is an estimated optimal path or not.
\end{itemize}

\subsubsection{ILP optimization problem}
\label{ss:MILP}
\noindent The optimization problem is presented in Eqs.~\ref{eq:optmprob_general}a--e.\\
\vspace{-0.8cm}
\begin{strip}
\begin{subequations}
\label{eq:optmprob_general}
\begin{align}
\max_{\hat{\mathbf{P}}, \mathbf{T}_{\mathtt{R}}} \;\; &  w_0 \cdot \Bigg(\sum_{k=1}^{|\mathcal{H}^{\mathcal{I}'}|} \hat{\mathbf{P}}_{(k)}\Bigg) - \sum_{m=1}^{|\mathtt{T}|}\sum_{n=1}^{|\mathtt{R}|} \mathbf{T}_{\mathtt{R}\,(m,\,n)} \label{eq:objfunc}\\
\textrm{s.t.} \;\; 
& \sum_{k=1}^{|\mathtt{R}|} \mathbf{H}_{\mathtt{R}{\,(m,\,k)}} \cdot \hat{\mathbf{P}}_{(m)} \cdot \mathbf{H}_{\mathtt{T}{\,(m,\,n)}} \leq 
\Big( \big( \mathbf{T}_{\mathtt{R}} \times \mathbf{H}_{\mathtt{R}}^{\top} \big) \cdot \mathbf{H}_{\mathtt{T}}^{\top} \Big)_{(n,m)}, \;\; \forall m \in \{1, ..., |\mathcal{H}^{\mathcal{I}'}|\}, \; \forall n \in \{1, ..., |\mathtt{T}|\} \label{eq:constraint1} \\
& \Big(\hat{\mathbf{P}} \times \mathbf{H}_{\mathtt{C}}\Big)_{(k)}  \leq 1, \;\;\;\; \forall k \in \{1, ..., |\mathtt{C}|\} \label{eq:constraint2}\\
& \sum_{k=1}^{|\mathtt{T}|} \mathbf{T}_{\mathtt{R}\,(k,\,m)} \leq 1, \;\;\;\;\; \forall m \in \{1, ..., |\mathtt{R}|\} \label{eq:constraint3}\\
& \; \hat{\mathbf{P}}_{(k)} \in \{0,1\}, \;\; \forall k \in \{1, ..., |\mathcal{H}^{\mathcal{I}'}|\}, \;\;\;\; \mathbf{T}_{\mathtt{R}\,(m,\,n)}, \in \{0,1\} \;\; \forall m \in \{1, ..., |\mathtt{T}|\}, \; \forall n \in \{1, ..., |\mathtt{R}|\}.
\label{eq:constraint4}
\end{align}
\end{subequations}
\end{strip}
\begin{itemize}
    \item[] \hspace{-0.7cm} We explain the different equations of the ILP of Eqs.~\ref{eq:optmprob_general}a--e:
    \item Equation \ref{eq:objfunc} aims to maximize the number of customers connected to their respective terminal elements by maximizing the number of estimated paths, $h \in \hat{\mathbf{P}}$, with value $1$. Additionally, a penalty term on the terminal association matrix $\mathbf{T}_{\mathtt{R}}$ is included, proportional to the number of elements assigned to each terminal element, to prevent unnecessary assignments.
    
    \item The constraint in Eq. \ref{eq:constraint1} verifies the validity of the paths included in the matrix $\hat{\mathbf{P}}$.
    It ensures that for every estimated path, $h \in \hat{\mathcal{P}}$, all its elements $e \in h$ are assigned to the same terminal element, $t \in \mathtt{T}$.
    
    \item The constraint in Eq. \ref{eq:constraint2} ensures that each customer, $c \in \mathtt{C}$, has at most one estimated path in $\hat{\mathbf{P}}$ with a value equal to $1$. This constraint is designed to prevent a customer from being associated with more than one estimated path.

    \item Equation \ref{eq:constraint3} guarantees that each element $r \in \mathtt{R}$ is associated to at most one specific terminal element $t \in \mathtt{T}$. Therefore, this constraint is designed to prevent the same elements from being associated with multiple terminal elements.
\end{itemize}

\subsection{Diagnostic function}
The set of estimated paths, $\hat{\mathcal{P}}$, identified is validated with a $Diagnostic()$ function. This function serves as a critical tool for assessing the validity of the solution. The $Diagnostic()$ function evaluates the found paths by checking any possible discrepancy from reality.\\
The diagnostic function outputs a list of issues encountered in the solution proposed, if any. If some issues are identified, it is possible to adjust the transformation functions to address these issues. Subsequently, the methodology is rerun to incorporate these changes until the results are acceptable.

\subsection{Solution of the TPI problem}
If no issues are identified or the DSO considers the solution to be acceptable, then the set $\hat{\mathcal{P}}$ is regarded as the final solution to the TPI problem.


\section{Academic example}
 \label{accexample}
To illustrate the methodology used to solve the TPI problem, an academic example of a power distribution network with a limited number of elements is considered.

\subsection{Raw information}
We assume the set of raw information available to the DSO, denoted as $\mathcal{I}$, is provided by some pieces of information:
\begin{enumerate}
    \item[$i_1$:] Documents containing the DSO's list of elements, their coordinates, and their types. However, some elements may be missing, and some coordinates may be inaccurate.
    \item[$i_2$:] MV/LV transformer cabin manuals provide instructions on cabin setup. This includes details about the organization of connection boards and how feeder lines connect to specific terminal elements or feeder terminal junctions on these boards.
    \item[$i_3$:] Customers' information about to which feeder terminal junction in the MV/LV transformer they are connected.
    \item[$i_4$:] DSO's general practice is to reduce energy losses by minimizing the total path length of each customer.
    \item[$i_5$:] Information that LV networks are operated radially.
\end{enumerate}

\subsection{Transformation functions}
The set $\mathcal{F}$ enumerates the transformation functions. Given that there are five pieces of raw information, the number of transformation functions is also five ($|\mathcal{F}|=|\mathcal{I}|=5$).

\subsection{Well-defined information}
\noindent The set of well-defined information is given by:
\begin{enumerate}
    \item[$i'_1$]$= f_1(i_1)$: Set of elements, $\mathcal{E}$, and their attributes, $\mathcal{A}$ are known.
    
    \item[$i'_2$]$= f_2(i_2)$: The elements of two types $junction$ 
    and $transformer$ are directly connected with no intermediate element. Since each feeder terminal junction has a known associated transformer, we can simplify the path identification problem. Instead of needing to track the path from the customer to the transformer, we only need to consider the path from the customer to the MV/LV transformer of the feeder junction terminal.
    
    \item[$i'_3$]$= f_3(i_3)$: For each element of type $customer$, $e \in \mathcal{E}$, its attribute $junction$, $c.junction$, where it is connected to is known.
    
    \item[$i'_4$]$= f_4(i_4)$: An element can be connected to another only if its distance is less than a given distance $D$. Moreover, the total length of a path must not exceed a maximum length $L$.
    
    \item[$i'_5$]$= f_5(i_5)$: Since the network has a radial structure and does not have loops, each element can only appear once in each path.

\end{enumerate}

\noindent The set of elements known to the DSO is given by:
\begin{align}
    \mathcal{E} = \{e_1, ..., e_{18}\}.
\end{align}
\noindent The set of attributes, $\mathcal{A}$, is given by:
\begin{align}
    \mathcal{A} = \{type, \; coordinate, \; junction\}.
\end{align}

\begin{figure}[h!]
\centering
\begin{subfigure}[b]{0.41\textwidth}
        \centering
        \includegraphics[width=\textwidth]{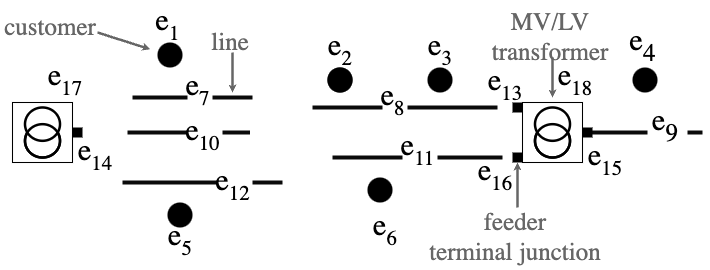}
        \caption{Academic network considered.}
        \label{fig:example1}
\end{subfigure}
\hfill
\begin{subfigure}[b]{0.41\textwidth}
        \centering
        \includegraphics[width=\textwidth]{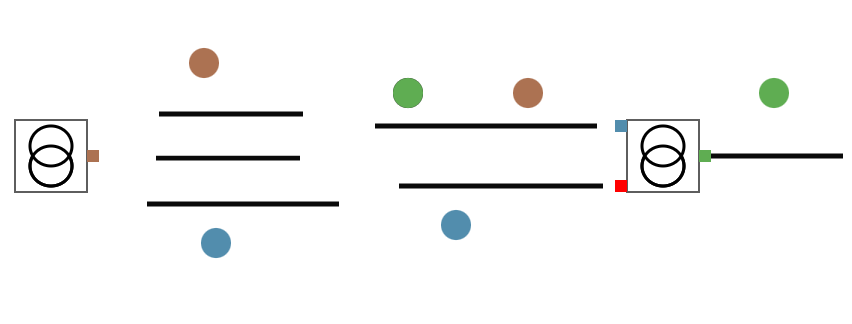}
        \caption{This visualization represents the assignment of customers to feeder terminal junctions within the network. Customers and feeder terminal junctions sharing the same color are supposed to be connected together.}
        \label{fig:example2}
\end{subfigure}
\hfill
\begin{subfigure}[b]{0.41\textwidth}
        \centering
        \includegraphics[width=\textwidth]{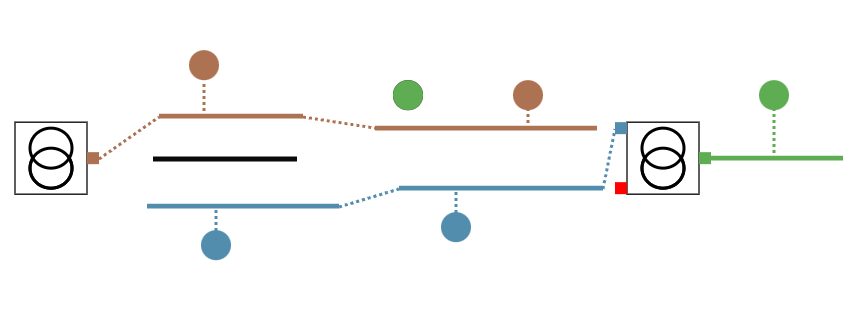}
        \caption{Results after the customer paths identification.}
        \label{fig:example3}
\end{subfigure}
\caption{Illustration of the process of identifying customer paths in the academic network considered.}
\end{figure}

\noindent The set of types, $\mathcal{T}$, is given by: $customer$ (elements $e_1$ to $e_6$), $line$ ($e_{6}$ to $e_{13}$), $junction$ ($e_{13}$ to $e_{16}$) and $transformer$ ($e_{17}$ and $e_{18}$) as shown in Fig. \ref{fig:example1}. 

\subsection{Hypothetical paths compatible with the well-defined information} 
The first two pieces of information are used to generate an initial set of hypothetical paths, $\mathcal{H}^{i'_1, i'_2}$.
Therefore, the size of the set of hypothetical paths, compatible with the first two pieces of the well-defined information, $i'_1$ and $i'_2$, is calculated using Eq. \ref{eq:size_h}:
\begin{align}
    |\mathcal{H}^{i'_1, i'_2}| = 61607.
\end{align}
The calculation is performed with the set of elements considered $\mathcal{E}$, 
the customer set $\mathtt{C}=Subset(\mathcal{E},customer)$, 
the terminal points set $\mathtt{T}=Subset(\mathcal{E},junction)$ and
the remaining elements set $\mathtt{R}=\mathcal{E}-\mathtt{C}-\mathtt{T} - Subset(\mathcal{E}, transformer)$.\\




\noindent Each remaining piece of well-defined information is used to exclude the paths in $\mathcal{H}^{i'_1, i'_2}$ that are not compatible with the well-defined information. Therefore, considering all the remaining pieces of well-defined information, $i'_3$ to $i'_5$, the total number of hypothetical paths compatible with the well-defined information is $|\mathcal{H}^{\mathcal{I}'}|=10$. \\
Detailed information about the elements in each hypothetical path can be found in the online repository (see Section \ref{repo}).

\subsection{Results paths estimation}
\noindent The matrices $\mathbf{H}_{\mathtt{C}}$, $\mathbf{H}_{\mathtt{R}}$, $\mathbf{H}_{\mathtt{T}}$, $\hat{\mathbf{P}}$ and $\mathbf{T}_{\mathtt{R}}$ are used in the optimization problem as shown in Eqs.~\ref{eq:optmprob_general}a--e.\\
In particular, the matrices $\mathbf{H}_{\mathtt{C}}$, $\mathbf{H}_{\mathtt{R}}$ and $\mathbf{H}_{\mathtt{T}}$ are used as the parameters of the ILP problem, while $\hat{\mathbf{P}}$ and $\mathbf{T}_{\mathtt{R}}$ are the decision variables and represent a proposed solution to the optimization problem.\\
The matrix $\hat{\mathbf{P}}$ contains the estimated paths are considered as an estimation of the real paths; while the matrix $\mathbf{T}_{\mathtt{R}}$ represents which elements belong to each feeder terminal junction.\\

\noindent The estimated optimal paths for this academic network are:
\begin{align}
    \hat{\mathbf{P}} =
    \begin{blockarray}{cccccccccc}
      h_1 & h_2 & h_3 & h_4 & h_{5} & h_{6} & h_{7} & h_{8} & h_{9} & h_{10} \\
    \begin{block}{(cccccccccc)}
      1 & 0 & 0 & 0 & 0 & 0 & 1 & 1 & 1 & 1 \\
    \end{block}
  \end{blockarray}
\end{align}

\noindent The matrix $\mathbf{T}_{\mathtt{R}}$ is given by:
\begin{equation}
\label{eq:Jsol}
    \mathbf{T}_{\mathtt{R}}   = 
        \begin{blockarray}{ccccccc}
        & e_7 & e_8 & e_{9} & e_{10} & e_{11} & e_{12} \\
        \begin{block}{c(cccccc)}
        e_{13} \;\; & 0 & 0 & 0 & 0 & 1 & 1 \\
        e_{14} \;\; & 1 & 1 & 0 & 0 & 1 & 1 \\
        e_{15} \;\; & 0 & 0 & 1 & 0 & 0 & 0 \\
        e_{16} \;\; & 0 & 0 & 0 & 0 & 0 & 0 \\
        \end{block}
        \end{blockarray}
\end{equation}
\vspace{-1cm}
\subsection{Results explanation}
We will now examine the results obtained from applying the optimization method to the academic example. This analysis will clarify the meaning and role of each component in Eqs.~\ref{eq:optmprob_general}a--e.\\
\subsubsection{Maximization term} 
The maximization term, Eq. \ref{eq:objfunc}, contains two operations:
\begin{itemize}
    \item Maximizing estimated paths: the first term, $\sum_{k=1}^{|\mathcal{H}^{\mathcal{I}'}|} \hat{\mathbf{P}}_{(k)}$, focuses on maximizing the number of estimated optimal paths, therefore paths whose have a value of $1$. 
    \item Penalty for unnecessary assignments: the second term, $\sum_{m=1}^{|\mathtt{T}|}\sum_{n=1}^{|\mathtt{R}|} \mathbf{T}_{\mathtt{R}\,(m,\,n)}$, introduces a penalty term related to the terminal association matrix. This term discourages assigning elements to a feeder terminal junction when unnecessary.\\
\end{itemize}

\subsubsection{Validity path constraint} 
The left-hand side of the constraint term in Eq. \ref{eq:constraint1} is calculated in three operations:
\begin{enumerate}[label=(\roman*)]
    \item $\sum_{k=1}^{|\mathtt{R}|} \mathbf{H}_{\mathtt{R}{\,(\bullet,\,k)}}$. This operation counts the number of elements present in each hypothetical path, $h \in \mathcal{H}^{\mathcal{I}'}$. The resulting matrix, of dimensions $|\mathcal{H}^{\mathcal{I}'}| \times 1$, is presented in transposed form for space efficiency as follows:
    \begin{align}
    \label{eq:constraint1_Li}
        \begin{blockarray}{cccccccccc}
          h_1 & h_2 & h_3 & h_4 & h_{5} & h_{6} & h_{7} & h_{8} & h_{9} & h_{10} \\
        \begin{block}{(cccccccccc)}
          1 & 2 & 2 & 2 & 3 & 3 & 2 & 1 & 2 & 1 \\
        \end{block}
      \end{blockarray}
    \end{align}
    \item The result, Eq. \ref{eq:constraint1_Li}, is multiplied element-wise by $\hat{\mathbf{P}}$. This operation assures the verification of the constraint only for the estimated optimal paths.
    The resulting matrix, still of dimensions $|\mathcal{H}^{\mathcal{I}'}| \times 1$, is transposed and presented as follows:
    \begin{align}
    \label{eq:constraint1_Lii}
        \begin{blockarray}{cccccccccc}
          h_1 & h_2 & h_3 & h_4 & h_{5} & h_{6} & h_{7} & h_{8} & h_{9} & h_{10} \\
        \begin{block}{(cccccccccc)}
          1 & 0 & 0 & 0 & 0 & 0 & 2 & 1 & 2 & 1 \\
        \end{block}
      \end{blockarray}
    \end{align}

    \item Finally, the result, Eq. \ref{eq:constraint1_Lii} is multiplied by the matrix $\mathbf{H}^{\top}$.
    The result of this operation still has dimensions $|\mathcal{H}^{\mathcal{I}'}| \times |\mathtt{T}|$ and the transposed form is shown as follows:
    \begin{equation}
    \label{eq:constraint1_Liii}
        \begin{blockarray}{ccccccccccc}
        & h_1 & h_2 & h_3 & h_4 & h_{5} & h_{6} & h_{7} & h_{8} & h_{9} & h_{10} \\
        \begin{block}{c(cccccccccc)}
        e_{13} \;\; & 0 & 0 & 0 & 0 & 0 & 0 & 0 & 0 & 2 & 1 \\
        e_{14} \;\; & 1 & 0 & 0 & 0 & 0 & 0 & 2 & 0 & 0 & 0 \\
        e_{15} \;\; & 0 & 0 & 0 & 0 & 0 & 0 & 0 & 1 & 0 & 0 \\
        e_{16} \;\; & 0 & 0 & 0 & 0 & 0 & 0 & 0 & 0 & 0 & 0 \\
        \end{block}
        \end{blockarray}
    \end{equation}
    The element-wise multiplication of matrices in Eq. \ref{eq:constraint1_Lii} and $\mathbf{H}^{\top}$ is handled through a process known as broadcasting. This technique allows matrices of different dimensions to be multiplied element-wise without explicitly resizing them.
    Broadcasting follows these key principles: matrices do not need to have identical dimensions for element-wise operations, and the resulting matrix adopts the dimensions of the input matrix with the largest dimensions. \\
    Formally, given a matrix $\mathbf{A}$ of dimensions $M \times 1$ and a matrix $\mathbf{B}$ of dimensions $M \times N$, with $M,N \in \mathbb{N}^+$ and the result of the product $\mathbf{A} \cdot \mathbf{B}$ is a matrix $\mathbf{C}$ of dimensions $M \times N$, where each element is given by:
    \begin{align}
        \mathbf{C}_{(m,n)} = \mathbf{A}_{(m,1)} \cdot \mathbf{B}_{(m,n)}, \;\; \forall m \in M, \; \forall n \in N.
    \end{align}
    Therefore, the resulting matrix in Eq. \ref{eq:constraint1_Liii} has dimensions $|\mathcal{H}^{\mathcal{I}'}| \times |\mathtt{T}|$.\\
\end{enumerate}

\noindent Similarly, the right-hand side of the constraint term in Eq. \ref{eq:constraint1} is calculated in two operations:
\begin{enumerate}[label=(\roman*)]
    \item the matrix multiplication of $\mathbf{T}_{\mathtt{R}}$ and $\mathbf{H}_{\mathtt{R}}^{\top}$. This operation calculates the number of elements that are present at the same time in a feeder terminal junction $t \in \mathtt{T}$ and in a path $h \in \mathcal{H}^{\mathcal{I}'}$. The resulting matrix has dimensions $|\mathtt{T}| \times |\mathcal{H}^{\mathcal{I}'}|$ and is given as follows:
    \begin{equation}
    \label{eq:constraint1_Ri}
        \begin{blockarray}{ccccccccccc}
        & h_1 & h_2 & h_3 & h_4 & h_{5} & h_{6} & h_{7} & h_{8} & h_{9} & h_{10} \\
        \begin{block}{c(cccccccccc)}
        e_{13} \;\; & 0 & 0 & 1 & 0 & 0 & 1 & 0 & 0 & 2 & 1 \\
        e_{14} \;\; & 1 & 1 & 1 & 1 & 2 & 1 & 2 & 0 & 0 & 0 \\
        e_{15} \;\; & 0 & 0 & 0 & 0 & 0 & 0 & 0 & 1 & 0 & 0 \\
        e_{16} \;\; & 0 & 0 & 0 & 0 & 0 & 0 & 0 & 0 & 0 & 0 \\
        \end{block}
        \end{blockarray}
    \end{equation}
    For example, considering the value at position $(e_{14},h_7)$ of the matrix in Eq. \ref{eq:constraint1_Ri}. The value is $2$ since the same elements $e_7$ and $e_{8}$ are at the same time assigned to the feeder terminal junction $e_{14}$ (second row of the junction association matrix in Eq. \ref{eq:Jsol}) and are present in the path $h_7$ (hypothetical path $h_7$ composed by the elements: $h_7=(e_{3}$, $e_{7}$, $e_{8}$, $e_{14})$).

    \item Similarly, for the left-hand side, the result of Eq. \ref{eq:constraint1_Ri} is multiplied by the matrix $\mathbf{H}_{\mathtt{T}}$. 
    This assures that the comparison on both sides of the inequality is performed only on the elements that are assigned to the same feeder terminal junction.
    The result of this operation has dimensions $|\mathtt{T}| \times |\mathcal{H}^{\mathcal{I}'}|$ and is expressed as follows:
    \begin{equation}
    \label{eq:constraint1_Rii}
        \begin{blockarray}{ccccccccccc}
        & h_1 & h_2 & h_3 & h_4 & h_{5} & h_{6} & h_{7} & h_{8} & h_{9} & h_{10} \\
        \begin{block}{c(cccccccccc)}
        e_{13} \;\; & 0 & 0 & 0 & 0 & 0 & 0 & 0 & 0 & 2 & 1 \\
        e_{14} \;\; & 1 & 0 & 0 & 0 & 0 & 0 & 2 & 0 & 0 & 0 \\
        e_{15} \;\; & 0 & 0 & 0 & 0 & 0 & 0 & 0 & 1 & 0 & 0 \\
        e_{16} \;\; & 0 & 0 & 0 & 0 & 0 & 0 & 0 & 0 & 0 & 0 \\
        \end{block}
        \end{blockarray}
    \end{equation}
\end{enumerate}
Finally, by comparing Eq. \ref{eq:constraint1_Liii} and Eq. \ref{eq:constraint1_Rii}, the condition Eq. \ref{eq:constraint1_Liii} $\leq$ Eq. \ref{eq:constraint1_Rii} is verified for each correspondent value of the matrices.\\

\noindent In general, the constraint Eq. \ref{eq:constraint1} assures that all the elements of an estimated path belong to the same feeder terminal junction. 
This is guaranteed when the number of elements in each estimated path is less than or equal to the total number of elements in the corresponding feeder terminal junction. \\


\subsubsection{Unique estimated path for customer constraint}
The constraint in Eq. \ref{eq:constraint2} is given as follows:
\begin{align}
\label{eq:result_expl_constraint2}
    & \Big(\hat{\mathbf{P}} \times \mathbf{H}_{\mathtt{C}}\Big) = 
        \begin{blockarray}{cccccc}
            e_1 & \! e_2 & \! e_{3} & \! e_{4} & \! e_{5} & \! e_{6} \\
            \begin{block}{(cccccc)}
            1 & \! 0 & \! 1 & \! 1 & \! 1 & \! 1 \\
            \end{block}
        \end{blockarray} \\
    & \Big(\hat{\mathbf{P}} \times \mathbf{H}_{\mathtt{C}}\Big)_{(k)} \leq 1, \;\;\;\; \forall k \in \{1, ..., |\mathtt{C}|\}
\end{align}
Equation \ref{eq:result_expl_constraint2} is a matrix of dimensions $1 \times |\mathtt{C}|$ and in this academic example, all the values are $1$ except for $e_{2}$ whose value is $0$ since no path was identified for that customer.\\

\subsubsection{Elements assigned to a unique feeder terminal junction constraint}
The constraint in Eq. \ref{eq:constraint3} is given as follows:
\begin{align}
\label{eq:result_expl_constraint3}
    & \sum_{k=1}^{|\mathtt{T}|} \mathbf{T}_{\mathtt{R}\,(k,\bullet)} = 
        \begin{blockarray}{cccccc}
            e_7 & \! e_8 & \! e_{9} & \! e_{10} & \! e_{11} & \! e_{12} \\
            \begin{block}{(cccccc)}
            1 & \! 1 & \! 1 & \! 0 & \! 1 & \! 1 \\
            \end{block}
        \end{blockarray} \\
    & \sum_{k=1}^{|\mathtt{T}|} \mathbf{T}_{\mathtt{R}\,(k,m)} \leq 1, \;\;\;\;\; \forall m \in \{1, ..., |\mathtt{R}|\} 
\end{align}
Equation \ref{eq:result_expl_constraint3} is a matrix of dimensions $1 \times |\mathtt{R}|$ and in this academic example, all the values are $1$ except for $e_{10}$ whose value is $0$ since the element $e_{10}$ is not used by any path, and therefore it is not assigned to any feeder terminal junction.

\subsection{Diagnostic function}
The proposed solution, the matrices $\hat{\mathbf{P}}$ and $\mathbf{T}_{\mathtt{R}}$ are then evaluated using the $Diagnostic()$ function to detect any possible issues. \\

\noindent For example, the solution proposed does not find the paths for each customer, since it is not possible to find a path for customer $e_2$ given the data available to the DSO. This issue is reported to the DSO, who can perform some further analysis on the case and understand why no path was found. \\
A possible reason is an error in the data and in reality that the customer belongs to another feeder terminal junction or a missing line in the DSO data. In such a particular case, a possible solution could be to assign another feeder terminal junction to the customer, depending on the (k-)closest customer(s).

\section{Belgian network} \label{casestudy} 
We apply our methodology to a real Belgian LV network characterized by incomplete GIS data and missing customer connections to the network. For this use case, we assume that the set of raw information, $\mathcal{I}$, the transformation functions, $\mathcal{F}$, and the well-defined information, $\mathcal{I}'$, are the same as those described in the academic example in Section \ref{accexample}.

\noindent For this real case, the set of elements known to the Belgian DSO is given by:
\begin{align}
    \mathcal{E} = \{e_1, ..., e_{1089}\}.
\end{align}
\noindent The set of attributes, $\mathcal{A}$, is given by:
\begin{align}
    \mathcal{A} = \{type, \; coordinate, \; junction\}.
\end{align}

\noindent The numbers of elements for the different types are given in Table \ref{tab:my-table} given below.:
\begin{table}[H]
\centering
\caption{Distribution Network Element Counts}
\label{tab:my-table}
\resizebox{0.8\columnwidth}{!}{%
\begin{tabular}{|c|c|}
\hline
\textbf{Subsets}          & \textbf{Number of elements} \\ \hline
$Subset(\mathcal{E}, customer)$    & 526                           \\ \hline
$Subset(\mathcal{E}, line)$        & 441                           \\ \hline
$Subset(\mathcal{E}, junction)$    & 96                           \\ \hline
$Subset(\mathcal{E}, transformer)$ & 26                           \\ \hline
\end{tabular}%
}
\end{table}

\subsection{Hypothetical paths compatible with the well-defined information}
The total number of hypothetical paths in the set $\mathcal{H}$ is generally determined using Eq. \ref{eq:size_h}. Following the methodology $\mathfrak{M}()$, the hypothetical paths not compatible with the well-defined information are excluded, keeping only those compatible with each piece of information, obtaining the set $\mathcal{H}^{I'}$. \\
For the real Belgian network considered, given the large number of elements in the network, it is impractical to explicitly represent the set $\mathcal{H}$ as a set of all the possible hypothetical paths.\\
\noindent For this reason, to efficiently identify the hypothetical paths that are compatible with the well-defined information, we develop a strategy to construct the set $\mathcal{H}^{\mathcal{I}'}$, or an approximation of it, without relying on the explicit enumeration of set $\mathcal{H}$.
The strategy is to use an $A^*()$ pathfinding algorithm (\cite{Astar}). The advantage of the $A^*()$ algorithm is to use a heuristic search approach to efficiently explore the possible hypothetical path space and to identify the hypothetical paths without exhaustively examining all possibilities.
The implementation of the $A^*()$ function is available on the online repository (see Section \ref{repo}). Below, we present the general concept. \\

\noindent 
The $A^*()$ function takes as input the set of network elements $\mathcal{E}$, the set of customers $\mathtt{C}$ for whom we want to identify hypothetical paths, the maximum number of hypothetical paths $N$ and some specific conditions (for example the maximum connection distance $D$ and the maximum length path $L$ as specified by the well-defined information $i'_3$).
For each customer, $c \in \mathtt{C}$, the $A^*()$ function explores its possible connections, using the function in Eq. \ref{eq:connections}, up to its feeder terminal junction, $c.junction$.
The $A^*()$ function then outputs a set of hypothetical paths, $\mathcal{H}^*$, that is consistent with the well-defined information and that approximate the set $\mathcal{H}^{\mathcal{I}'}$. \\

\noindent After the execution of the $A^*()$ algorithm, considering a maximum number of paths for each customer of $N=5$, the total number of hypothetical paths is $|\mathcal{H}^{\mathcal{I}'}| = 2348$. The value of $N$ is selected as a balance between computational efficiency and the likelihood of identifying an optimal solution.

\subsection{Matrix generation}
The set $\mathcal{H}^{\mathcal{I}'}$ is decomposed into the three sub-matrices, 
$\mathbf{H}_{\mathtt{C}}$, 
$\mathbf{H}_{\mathtt{R}}$ and 
$\mathbf{H}_{\mathtt{T}}$, as explained in Section \ref{ss:mat_dec}.\\
These matrices are used as input parameters of the optimization problem.\\

\noindent The matrices $\hat{\mathbf{P}}$ and $\mathbf{T}_{\mathtt{R}}$ are used as output decision variables of the optimization algorithm.

\subsection{Optimization problem results}
After executing the optimization algorithm, a solution is obtained in the form of matrices $\hat{\mathbf{P}}$ and $\mathbf{T}_{\mathtt{R}}$.\\
Displaying the matrices directly might be challenging for visualization and comprehension. Therefore, we summarize and present the results in a more comprehensible way. \\
\noindent Therefore, given the matrix, $\hat{\mathbf{P}}$ it is possible to represent graphically the connections among the elements of the estimated paths. In particular, each element of an estimated optimal path is assigned a specific color, with black reserved for unassigned elements.\\
The color that is assigned to each element depends on the matrix $\mathbf{T}_{\mathtt{R}}$. In particular, the same color is assigned to the elements that belong to the same feeder terminal junction.

\begin{figure}[h!]
\centering
\begin{subfigure}[b]{0.47\textwidth}
        \centering
        \includegraphics[width=\textwidth]{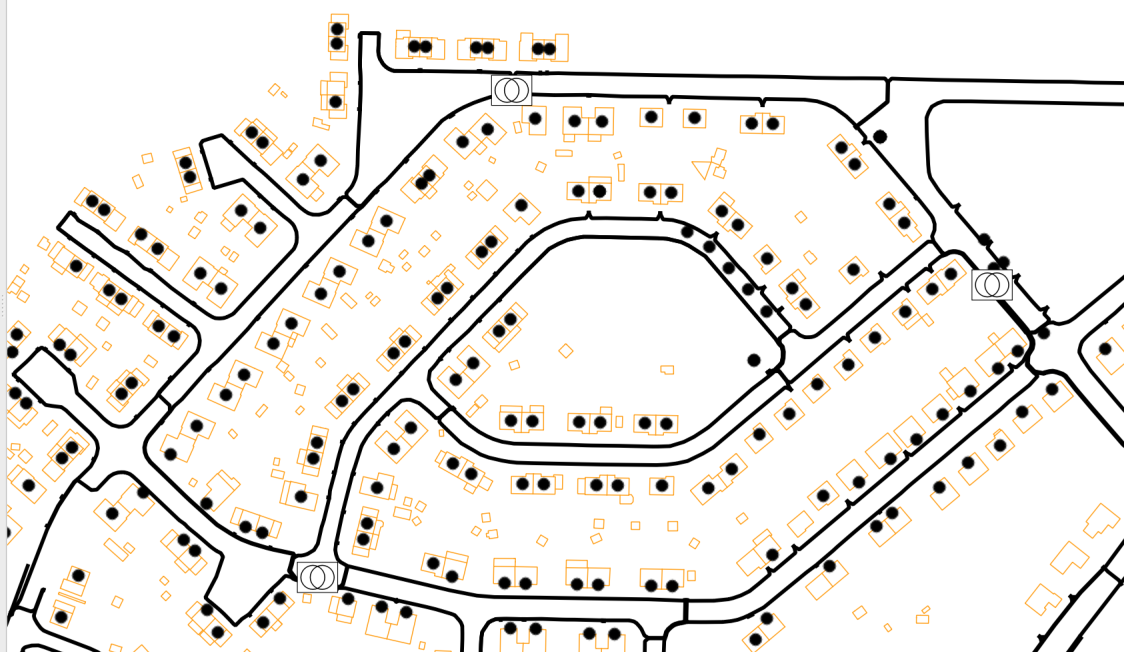}
        \caption{Before applying the methodology.}
        \label{fig:realnetowrk}
\end{subfigure}
\hfill
\begin{subfigure}[b]{0.47\textwidth}
        \centering
        \includegraphics[width=\textwidth]{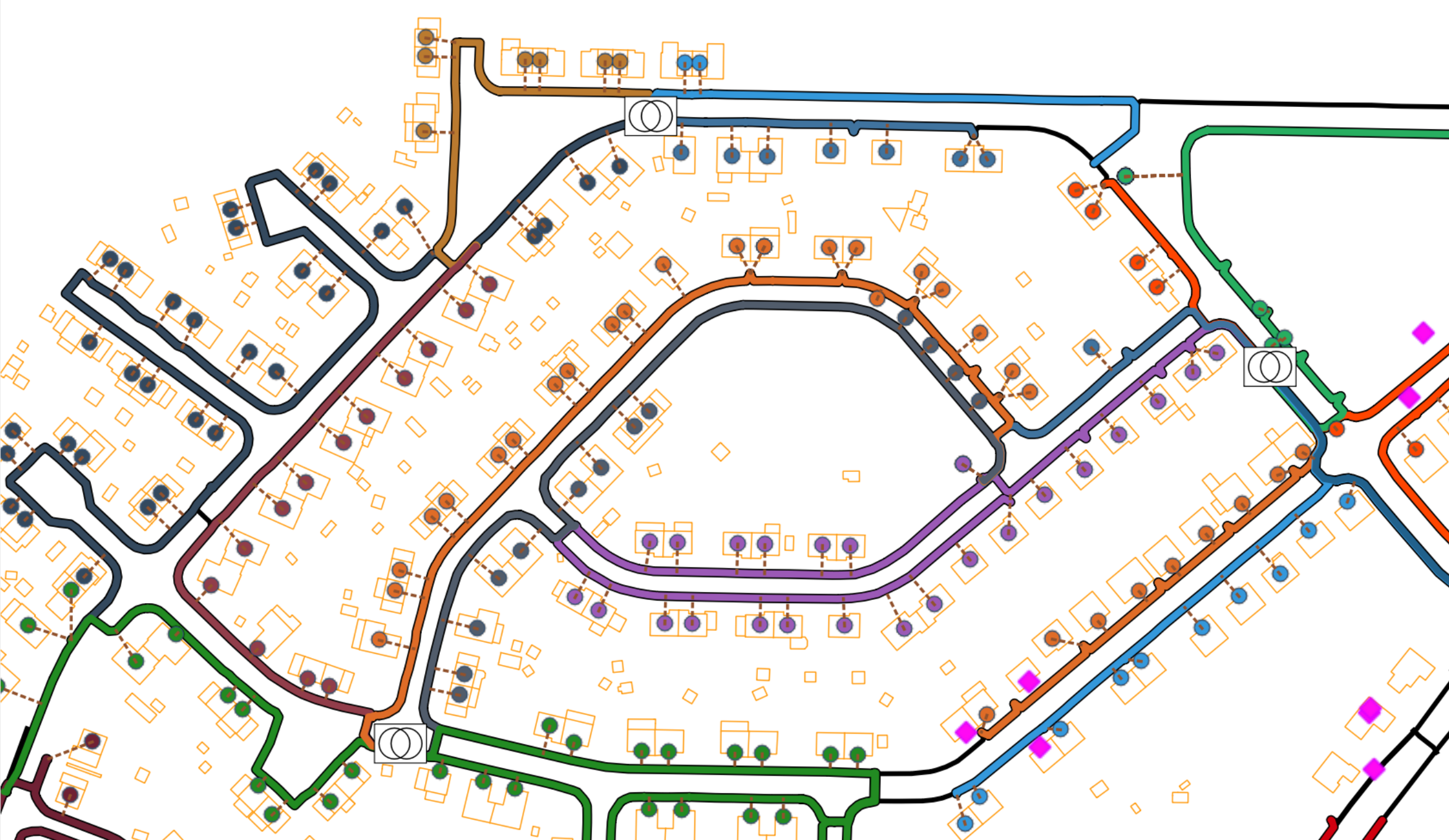}
        \caption{After applying the methodology.}
        \label{fig:solutionrealnetowrk}
\end{subfigure}
\caption{Part of the real network considered.}
\label{fig:}
\end{figure}

\noindent Figure \ref{fig:realnetowrk} shows a part of the Belgian network considered with three MV/LV transformers, some lines, and customers.
In Fig. \ref{fig:solutionrealnetowrk} it is possible to see the network after the identification of the paths: colors have been assigned to customers and lines belonging to the same feeder terminal junction. Dashed lines represent the connections between the elements, for example, customer-line connections.

\subsection{Diagnostic function}
Moreover, few purple squares are present. These are customers for which no path was identified and these are discovered with the $Diagnostic()$ function.\\
Possible reasons for failing to identify a path include incorrect or missing data for the feeder terminal junction associated with the customer, or the inability to assign a line to its corresponding feeder terminal junction. \\
These issues are reported to the DSO for further analysis.
\vspace{-0.2cm}

\section{Conclusion} \label{conclusion}
This paper introduces an optimization algorithm aimed at addressing the topological path identification (TPI) problem in power distribution networks. By leveraging information provided by distribution system operators (DSOs), the methodology constructs a set of hypothetical paths. This set is reduced by excluding the paths not compatible with the well-defined information. After that, an integer linear programming (ILP) algorithm identifies the best approximation of the real paths by selecting the paths that maximize customer connectivity to the correct medium-voltage/low-voltage (MV/LV) transformer. 
Our approach is based only on static data, without the requirements of data coming from advanced metering infrastructure (AMI) like smart meters. This makes the methodology more general and applicable to a wide range of distribution networks, including those in regions with limited technological infrastructure. Additionally, it minimizes the cost and complexity associated with collecting and integrating data from dynamic sources.
The effectiveness of the proposed approach has been validated through demonstrations on both academic examples and real-world networks. 
This methodology is fundamental for constructing accurate digital twins of power distribution networks, ultimately aiding DSOs in network management and optimization.\\
Possible future work includes expanding the optimization algorithm to identify backup paths, using AMI sources such as smart meters and sensors to enhance the accuracy of the path identification, and improving the scalability and computational efficiency of the optimization algorithm.

\vspace{-0.2cm}
\section*{References}
\printbibliography[heading=none]

@inproceedings{TPI2024,
	AUTHOR = {Vassallo, Maurizio and Bahmanyar, Alireza and Duchesne, Laurine and Leerschool, Adrien and Gerard, Simon and Wehenkel, Thomas and Ernst, Damien},
	TITLE = {A Systematic Procedure for Topological Path Identification with Raw Data Transformation in Electrical Distribution Networks},
	YEAR = {2024},
	MONTH = {04},
	PUBLISHER = {CEEPE},
	BOOKTITLE = {Proceedings of International Conference on Energy, Electrical and Power Engineering},
	LOCATION = {Yangzhou, China},
doi = {10.1109/CEEPE62022.2024.10586328}
}

@inproceedings{NavarroEspinosa2015RECONSTRUCTIONOL,
  title={RECONSTRUCTION OF LOW VOLTAGE DISTRIBUTION NETWORKS: FROM {GIS} DATA TO POWER FLOW MODELS},
  author={Alejandro Navarro-Espinosa and Luis F. Ochoa},
  year={2015},
  url={https://api.semanticscholar.org/CorpusID:55344363}
}

@article{ZepuKnowledge2020,
author = {Zepu, Gao and Yongjian, Luo and Ziwei, Xu and Yilan, Yu and Lianmei, Zhang},
title = {Knowledge graph-based method for identifying topological structure of low-voltage distribution network},
journal = {The Journal of Engineering},
volume = {2020},
number = {12},
pages = {1177-1184},
doi = {10.1049/joe.2019.1319},
year = {2020}
}

@ARTICLE{WangPower2021,
AUTHOR={Wang, Changgang and An, Jun and Mu, Gang},
TITLE={Power System Network Topology Identification Based on Knowledge Graph and Graph Neural Network},
JOURNAL={Frontiers in Energy Research},
YEAR={2021},
DOI={10.3389/fenrg.2020.613331},      
}

@article{farajollahi_topology_2020,
	title = {Topology Identification in Distribution Systems Using Line Current Sensors: An {MILP} Approach},
	volume = {11},
	doi = {10.1109/TSG.2019.2933006},
	pages = {1159--1170},
	number = {2},
	journaltitle = {{IEEE} Transactions on Smart Grid},
	author = {Farajollahi, Mohammad and Shahsavari, Alireza and Mohsenian-Rad, Hamed},
	date = {2020-03},
}

@article{soltani_real-time_2022,
	title = {Real-Time Topology Detection and State Estimation in Distribution Systems Using Micro-{PMU} and Smart Meter Data},
	volume = {16},
	doi = {10.1109/JSYST.2022.3153449},
	pages = {3554--3565},
	number = {3},
	journaltitle = {{IEEE} Systems Journal},
	author = {Soltani, Zahra and Khorsand, Mojdeh},
	date = {2022-09},
}

@INPROCEEDINGS{RLVNLO2021,
  author={Marulli, Daniele and Mathieu, Sébastien and Benzerga, Amina and Sutera, Antonio and Ernst, Damien},
  booktitle={2021 IEEE PES Innovative Smart Grid Technologies Europe (ISGT Europe)}, 
  title={Reconstruction of Low-Voltage Networks with Limited Observability}, 
  year={2021},
  pages={5},
  doi={10.1109/ISGTEurope52324.2021.9640163}}

@ARTICLE{RNMLSDP2011,
  author={Mateo Domingo, Carlos and Gomez San Roman, Tomas and Sanchez-Miralles, Alvaro and Peco Gonzalez, Jesus Pascual and Candela Martinez, Antonio},
  journal={IEEE Transactions on Power Systems}, 
  title={A Reference Network Model for Large-Scale Distribution Planning With Automatic Street Map Generation}, 
  year={2011},
  volume={26},
  number={1},
  pages={190-197},
  doi={10.1109/TPWRS.2010.2052077}}

@inproceedings{Seack2014GENERATINGLV,
  title={GENERATING LOW VOLTAGE GRIDS ON THE BASIS OF PUBLIC AVAILABLE MAP DATA},
  author={Andre Seack and Jan Kays and Christian Rehtanz},
  booktitle={CIRED workshop 2014},
  url={https://api.semanticscholar.org/CorpusID:38279545}
}

@incollection{li2008distribution,
  author = {Li, X. and Feng, X. and Zeng, Z. and Xu, X. and Zhang, Y.},
  title = {Distribution feeder one-line diagrams automatic generation from geographic diagrams based on {GIS}},
  booktitle = {2008 Third International Conference on Electric Utility Deregulation and Restructuring and Power Technologies},
  pages = {2228--2232},
  publisher = {IEEE},
  year = {2008},
  doi={10.1109/DRPT.2008.4523781}
}

@inproceedings{wu2018distribution,
  author = {Wu, L. and Lin, Y. and Pang, W.},
  title = {Distribution Network Topology Modelling and Automatic Mapping Based on {CIM} and {GIS}},
  booktitle = {2018 IEEE 4th Information Technology and Mechatronics Engineering Conference (ITOEC)},
  pages = {5},
  organization = {IEEE},
  year = {2018},
  doi = {10.1109/ITOEC.2018.8740469}
}

@inproceedings{shang2019automatic,
  author = {Shang, L. and Hu, R. and Ci, H. and Zhang, W. and Ouyang, G.},
  title = {Automatic Generation Algorithm of Distribution Network Topology Map Based on {GIS} Drawing},
  booktitle = {IOP Conference Series: Earth and Environmental Science},
  volume = {384},
  number = {1},
  year = {2019},
doi = {10.1088/1755-1315/384/1/012231},
}

@inproceedings{yin2021research,
  author = {Yin, Z. and Luo, J. and Wang, L. and Ye, R.},
  title = {Research on Network Topology Analysis Method of Distribution Management Based on {GIS}},
  booktitle = {2021 International Conference on Intelligent Transportation, Big Data and Smart City (ICITBS)},
  pages = {145--148},
  organization = {IEEE},
  year = {2021},
doi = {10.1109/ICITBS53129.2021.00045}
}

@article{cui2022low,
  title={Low-voltage distribution network topology identification based on constrained least square and graph theory},
  author={Cui, Shijie and Zeng, Peng and Song, Chunhe and Wang, Zhongfeng and Li, Guangye},
  journal={Soft Computing},
  volume={26},
  number={17},
  pages={8509--8519},
  year={2022},
  publisher={Springer},
doi = {https://doi.org/10.1016/j.epsr.2022.108969}
}

@INPROCEEDINGS{wang2023generation,
  author={Wang, Shaowu and Chen, Hongzhong and Dai, Feng and He, Yu},
  booktitle={2023 International Conference on Computer Science and Automation Technology (CSAT)}, 
  title={The Automatic Generation Algorithm for Single Line Diagrams of Distribution Networks Based on the Layout Model of Branch and Main Lines}, 
  year={2023},
  volume={},
  number={},
  pages={636-640},
  doi={10.1109/CSAT61646.2023.00172}}

@ARTICLE{Yu2024identification,
author={Yu, Xiao and Zhao, Jian and Zhang, Haipeng and Wang, Xiaoyu and Bian, Xiaoyan},
journal={IEEE Transactions on Industrial Informatics}, 
title={Data-Driven Distributed Grid Topology Identification Using Backtracking {Jacobian} Matrix Approach}, 
year={2024},
volume={20},
number={2},
pages={1711-1720},
doi={10.1109/TII.2023.3280936}}

@article{li2022topology,
title = {Topology identification method for residential areas in low-voltage distribution networks based on unsupervised learning and graph theory},
journal = {Electric Power Systems Research},
volume = {215},
year = {2023},
doi = {https://doi.org/10.1016/j.epsr.2022.108969},
author = {Haifeng Li and Wenzhao Liang and Yuansheng Liang and Zhikeng Li and Gang Wang},
}

@inproceedings{benzerga2021low,
  title={Low-voltage network topology and impedance identification using smart meter measurements},
  author={Benzerga, Amina and Maruli, Daniele and Sutera, Antonio and Bahmanyar, Alireza and Mathieu, S{\'e}bastien and Ernst, Damien},
  booktitle={2021 IEEE Madrid PowerTech},
  pages={6},
  year={2021},
  organization={IEEE},
doi={10.1109/PowerTech46648.2021.9495093}
}

@inproceedings{morrell2018modelling,
author = {Morrell, Thomas and Venkataramanan, Venkatesh and Srivastava, Anurag and Bose, Anjan and Liu, Chen-Ching},
year = {2018},
month = {04},
pages = {9},
title = {Modeling of Electric Distribution Feeder Using Smart Meter Data},
doi = {10.1109/TDC.2018.8440540}
}

@INPROCEEDINGS{parikh2009transforming,
  author={Parikh, P. A. and Nielsen, T. D.},
  booktitle={2009 IEEE/PES Power Systems Conference and Exposition}, 
  title={Transforming traditional geographic information system to support smart distribution systems}, 
  year={2009},
  volume={},
  number={},
  pages={4},
  doi={10.1109/PSCE.2009.4839979}}

@INPROCEEDINGS{peppanen2016estimation,
  author={Peppanen, Jouni and Grijalva, Santiago and Reno, Matthew J. and Broderick, Robert J.},
  booktitle={2016 IEEE/PES Transmission and Distribution Conference and Exposition}, 
  title={Distribution system low-voltage circuit topology estimation using smart metering data}, 
  year={2016},
  volume={},
  number={},
  pages={5},
doi = {10.1109/TDC.2016.7519985}
}

@INPROCEEDINGS{Guzman2017identification,
author={Guzmán, Abdenago and Quirós-Tortós, Jairo and Valverde, Gustavo},
booktitle={2017 IEEE Manchester PowerTech}, 
title={Efficient connectivity identification of large-scale distribution network elements in {GIS}}, 
year={2017},
volume={},
number={},
pages={6},
doi={10.1109/PTC.2017.7981064}}

@ARTICLE{PTI2024,
  author={Pengwah, Abu Bakr and Gerdroodbari, Yasin Zabihinia and Razzaghi, Reza and Andrew, Lachlan L. H.},
  journal={IEEE Transactions on Power Delivery}, 
  title={Topology Identification of Distribution Networks With Partial Smart Meter Coverage}, 
  year={2024},
  volume={39},
  number={2},
  pages={992-1001},
  doi={10.1109/TPWRD.2024.3354292}}

@ARTICLE{Astar,
  author={Hart, Peter E. and Nilsson, Nils J. and Raphael, Bertram},
  journal={IEEE Transactions on Systems Science and Cybernetics}, 
  title={A Formal Basis for the Heuristic Determination of Minimum Cost Paths}, 
  year={1968},
  volume={4},
  number={2},
  pages={100-107},
  doi={10.1109/TSSC.1968.300136}}


\end{document}